\documentclass[epj]{svjour}
\usepackage{epsfig,epsf}
\begin{document}

\title{Resonant fluxon transmission through   impurities}
\author{ Yaroslav Zolotaryuk}
\institute{
 Bogolyubov Institute for Theoretical Physics,
National Academy of Sciences of Ukraine,
vul. Metrologichna 14B,
03680 Kyiv, Ukraine,
\email{yzolo@bitp.kiev.ua}
}
\date{Received: date / Revised version: date}
\abstract{ 
Fluxon transmission through several impurities of different 
strength and type (i.e., microshorts and microresistors), 
placed in a long Josephson junction is investigated.
Threshold pinning current on the impurities is computed
as a function of the distance between them, their amplitudes and
the dissipation parameter. It is shown that in the case
of consequently placed microshorts or microresistors, the threshold
pinning current exhibits a clear minimum as a function of the
distance between the impurities. In the case of a microresistor,
followed by a microshort, an opposite phenomenon is observed, namely
the threshold pinning current exhibits maximum as a function of
the distance between the impurities. 
}
\maketitle

\section{Introduction and the background}

The dynamics of magnetic flux propagation in a long Josephson
junction (LJJ) is a subject of increasing theoretical and 
practical interest \cite{barone82,u98pd}. Magnetic flux quantum
in a LJJ is a soliton (also known as {\it fluxon}) 
governed by the well-known sine-Gordon (SG) equation.
A convenient way to prepare a junction with the required 
properties is to install various inhomogeneities into it.
Up to now substantial work has been devoted to the study of the
fluxon motion in the LJJs with point-like impurities. The interaction
of a fluxon with a single impurity became a textbook 
example \cite{ms78pra}. 

On the other hand, the phenomenon of resonant tunneling of an
electron through a
double-well structure is well-known in quantum mechanics \cite{restun}. 
A natural question arises: what is an analog of the quantum-mechanical
resonant tunneling in the fluxon dynamics? Resonant soliton 
transmission has been investigated in detail for non-dissipative
systems \cite{ksckv92jpa,zkv94jpsj} and complex resonant 
behaviour has been reported. However, fluxon dynamics in a LJJ cannot
be considered without taking into account dissipative effects, 
which are a consequence of the normal electron tunneling across
the insulating barrier. As a result, transmission in a LJJ with 
constant bias and dissipation can yield only two scenarios: fluxon
transmission or fluxon pinning on the impurities. And, 
consequently, the transmission ratio
can attain only two values: zero or unity. Therefore the attention has to
be turned toward other characteristic quantities, especially the minimal
bias, necessary for the fluxon pinning on impurities.

The present paper aims to investigate fluxon transmission through
several (two or more) point-like impurities: microshorts, microresistors
or a combination of both. Of particular interest is dependence of the 
threshold pinning current on the distance between the impurities and their
amplitudes.

The paper is organized as follows. In Section  \ref{s2} we present
the model and the basic equations of motion.  In the
next section we describe the methods of the analysis of the
equations and motion and study the 
fluxon transmission through two microshorts, two microresistors
and a microshort and a microresistor as a function of their
amplitudes and distance between them. Discussion of the obtained
results and final remarks are given in Sec. \ref{s4}.

\section{The model}
\label{s2}

We consider the long Josephson junction (LJJ) subjected to the external 
time-independent bias. The main dynamical variable is the difference 
between the phases $\theta_2(x,t)-\theta_1(x,t)=\phi(x,t)$ of the
macroscopic wave functions of the the superconducting layers of the
junction. The time evolution of the phase difference is governed by
the perturbed sine-Gordon (SG) equation: 
\begin{eqnarray}
\nonumber
 &&\phi_{tt}-\phi_{xx}+ \sin \phi = \epsilon f [\phi, \phi_t; x]~, \\
 && \epsilon f [\phi, \phi_t; x]\doteq -\alpha \phi_t - 
         \gamma-\sin \phi \sum_{n=1}^{m} \mu_n \delta (x-a_m).
\label{1}
\end{eqnarray}
In this dimensionless equation spacial variable $x$ is normalized to 
the Josephson penetration depth $\lambda_J$, the temporal variable $t$
is normalized to the inverse 
Josephson plasma frequency $\omega_J^{-1}$\cite{parameters}. 
Here the bias current $\gamma$ is normalized to the critical Josephson 
current of the junction and the dimensionless parameter $\alpha$ 
describes dissipation. It is supposed that there are $N$ impurities 
in this junction, 
positioned at the points $x=a_n$, $n=1,2,\ldots,N$, 
$a_1\equiv 0<a_2<...<a_m$, with $\mu_n$ being ``strength'' or amplitude
of the $n$th impurity. The impurity is a microshort if $\mu_n>0$ and a 
microresistor if $\mu_n<0$.

\section{Fluxon transmission}
\label{s3}

A standard tool for analyzing the fluxon dynamics in Josephson 
junctions is the McLaughlin-Scott perturbation theory \cite{ms78pra}. 
Also, direct numerical 
integration\footnote{
In the numerical simulations, the space will be discretized 
as $x \to nh$, so that the continuous variable 
$\phi(x,t)\simeq \phi(nh,t)$ becomes a discrete set of 
variables $\phi_n(t)$, and the second space
derivative becomes 
$\phi_{xx}(x,t) \simeq [\phi_{n+1}(t)-2\phi_n(t)+\phi_{n-1}(t)]/h^2$.
The resulting set of the second order ODEs on $\phi_n(t)$
will be solved using the 4th order Runge-Kutta scheme. 
The delta function is approximated as 
$\delta(x)\simeq \delta_{n,0}/h$ where $\delta_{m,n}$ is 
Kronecker's $\delta$ symbol.
} 
of the perturbed SG equation 
(\ref{1}) will be performed to check the validity of the analytical  
approximation. We are going to solve the problem for the idealized case
of an infinite junction with free ends boundary conditions,
however, in actual simulation a sample with length that significantly
exceeds the fluxon size will be used.

\subsection{Perturbation theory and collective coordinates}
\label{s3.0}

Using the perturbation theory, one obtains in the first order the 
evolution equations for the fluxon
parameters, i.e., its center of mass $X$ and fluxon velocity $v$:
\begin{eqnarray}
\nonumber
&&\dot v = \frac{\pi \gamma}{4}(1-v^2)^{3/2}-\alpha v(1-v^2)+\\
&&        +\frac{1-v^2}{2}
\label{2}
 \sum_{n=1}^N \mu_n g [X-a(n-1),v]~,\\  
&&\dot X = v - \frac{v}{2} 
\label{3}
  \sum_{n=1}^N \mu_n [X-a(n-1)]g[X-a(n-1),v],\\
&&g(X,v)\doteq \frac{\tanh{(X/\sqrt{1-v^2})}}{\cosh^2{(X/\sqrt{1-v^2})}}~.
\nonumber
\end{eqnarray}
For the sake of simplicity in the following only equidistant impurities 
will be considered, i.e., $a_n \equiv a$, $n=1,2,\ldots,N$.
Also, only positive values of bias $\gamma$ will be considered.
The case of one impurity ($N=1$, $\mu_1\equiv \mu$) has been discussed
in detail in \cite{ms78pra,kmn88jetp}. There exist two characteristic values of 
the bias current, $\gamma_c \equiv 4\sqrt{3}\mu/(9\pi)$ and 
$\gamma_{thr}$, $\gamma_c > \gamma_{thr}$. If $\gamma > \gamma_c$, the 
pinning on the impurity is not possible and only one attractor that 
corresponds to fluxon propagation does exist. In the interval 
$\gamma_{thr}< \gamma < \gamma_c$ two attractors exist: one corresponds
to fluxon pinning on the microshort and another one to fluxon 
propagation. If $\gamma < \gamma_{thr}$, the only possible regime is 
fluxon pinning on the impurity. It has been 
shown \cite{ms78pra,kmn88jetp} that there exists a threshold value
of the dc bias, which can be approximated as
\begin{equation}
\gamma_{thr}=\frac{\alpha}{\pi}\sqrt{8 \mu + \mu^2}
~\left [1-2\alpha \ln 2 \right ]~.
\label{4a}
\end{equation}
In the case of one microresistor $\mu<0$, the threshold bias has been 
defined in \cite{kmn88jetp} as
\begin{equation}
\gamma_{thr}=2\frac{\alpha}{\pi} \left [2\sqrt{\sqrt{2|\mu|}\pi\alpha}
+\frac{9}{2}\alpha  \ln (\alpha \sqrt{2/|\mu|})\right ]~.
\label{4b}
\end{equation}
In the non-relativistic limit ($v \ll 1 $) the system (\ref{2})-(\ref{3})
can be rewritten
as a Newtonian second order ODE for the particle of mass $8$ 
(see Refs. \cite{ms78pra,kmn88jetp}):
\begin{eqnarray}
\label{6a}
&&8 \ddot X+ 8\alpha \dot X = - \frac{\partial U}{\partial X}~,\\
\label{6b} 
&& U(X)=-2 \pi \gamma X+U_0(X) \equiv -2 \pi \gamma X+\\
\nonumber
&&+2\sum_{n=1}^m \frac{\mu_n}{\cosh^2 {[X-a(n-1)]}}~,~~ \gamma \geq 0~.
\end{eqnarray}
In the case of strongly separated impurities 
($a \gg 1$), the potential $U(X)$ has $2N$ extrema [they approximately 
coincide with the fixed points $X=X_{2k}$, $k=1,2,\ldots,N$, of equations 
(\ref{2})-(\ref{3})] 
where each pair (a minimum and a maximum) is associated with a certain
impurity. If there is an impurity at $X=a(k-1)$,  
the minimum at $X=X_{2k-1}$ always comes 
before the maximum at $X=X_{2k}$. Microshorts are repelling
impurities, thus the fluxon that arrives from $X=-\infty$ 
decelerates when approaching it and accelerates 
after passing the impurity until the fluxon velocity reaches 
the equilibrium value  
\begin{equation}
v_\infty=\left [ 1+ \left( \frac{4\alpha}{\pi\gamma}\right )\right ]^{-1/2}~.
\label{7}
\end{equation}
Microresitors are
attractive impurities, and, as a result, the fluxon accelerates before 
approaching the microresistor
and slows down to the equilibrium velocity (\ref{7}) after being 
released from it. Decrease of the distances between the impurities
$a$ causes disappearance of some of the extrema via inverse pitchfork 
bifurcations. The systematic phase plane analysis of equations 
(\ref{2})-(\ref{3}) for the case of two 
microshorts has been performed in \cite{bp90pla} for the SG 
equation and in \cite{pb94pla} for the double SG
equation. In those papers the behaviour of the fixed points
of the system (\ref{2})-(\ref{3}) has been studied as a function
of the distance between them, $a$.

Our aim is to determine the threshold current 
$\gamma_{thr}=\gamma_{thr}(a,\{\mu_n \}_1^N;N)$ as a function of
the distance between impurities and their amplitudes. In the
case of one impurity, $\gamma_{thr}(\mu;1)$ obviously does not
depend on the distance $a$. It is described approximately by 
equations (\ref{4a})
and (\ref{4b}) for the microshort and microresistor, respectively.
Some general statement can be made before one proceeds
to specific cases. Two important limits should be mentioned. 
One case corresponds to impurities being separated
by the distance much larger the fluxon size. Then the transmission will be
governed by the fluxon interaction with each individual impurity. In 
the opposite limit ($a \rightarrow 0$) the power of all impurities 
adds up. The effect of the both limits on the threshold current can be 
written as follows
\begin{equation}
\label{6}
\gamma_{thr}(a,\{\mu_n \}_1^N;N) = \left \{ 
\begin{array}{c}
\gamma_{thr}\left (\sum_{n=1}^N \mu_n;1 \right ),~a \to 0, \\
\max_{1\le n\le N}[ \gamma_{thr}(\mu_n;1) ],~a \to \infty.
\end{array}
\right .
\end{equation}

In the subsections below the transmission through impurities of different 
polarities (e.g, $\mu_n<0$ and $\mu_n>0$) will be considered. 
It appears that for $N \ge 2$ the analytical treatment of equations 
(\ref{2})-(\ref{3}) is virtually not possible even in the
non-relativistic case, especially when none of the limits,
described by the equation (\ref{6}), hold. Therefore equations 
(\ref{2})-(\ref{3}) are going to be solved numerically.

\subsection{Transmission through two microshorts}
\label{s3.1}

\begin{figure}[htb]
\centerline{\epsfig{file=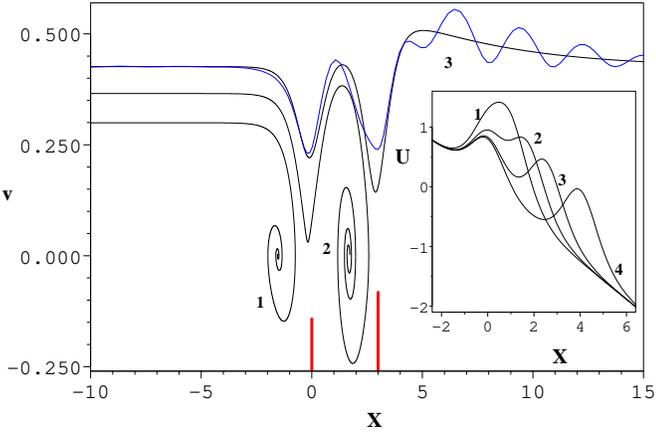,width=2.8in,angle=-90}}
\caption{Phase space trajectories of equation (\ref{2}) for $N=2$ 
microshorts with
$\mu_1=0.4$, $\mu_2=0.6$ and distance $a=3$ between them. Curve $1$
corresponds to $\gamma=0.04$, curve $2$ -  to $\gamma=0.05$ and
curve $3$ - to $\gamma=0.06$. Blue curve shows results of
direct integration of the equation (\ref{1}) at $\gamma=0.06$ 
the using Runge-Kutta method. The fluxon center of mass 
is defined as $X(t)=\int_{-\infty}^{+\infty} x \phi_x dx $ and
its velocity as $v(t)=dX(t)/dt$.
Dissipation for all cases is $\alpha=0.1$.
The inset shows the effective potential $U(X)$ [equation (\ref{6b})]
for $\mu_1=0.4,\mu_2=0.6$, $\gamma=0.05$, $a=1$ (curve 1),
$a=1.7$ (curve 2), $a=2.5$ (curve 3) and $a=4$ (curve 4).
Red vertical bars denote locations of the microshorts and the lengths
of the bars is proportional to the microshort amplitudes, $\mu_{1,2}$.
}
\label{fig1}
\end{figure}

Consider first the case of two microshorts ($N=2$, $\mu_{1,2}>0$). 
The problem is tackled in the following way. The fluxon approaches 
the system of two microshorts from $X=-\infty$ with the equilibrium 
velocity $v_\infty$, given by equation (\ref{7}). Evolution of the
system (\ref{2})-(\ref{3}) on the phase plane $(X,v)$ is shown 
in Figure \ref{fig1}. Depending on the strength
of the bias, three scenarios are possible: trapping on the first 
microshort (curve 1);
trapping on the second microshort, if the external bias is
a bit larger (curve 2); or transmission (curve 3).  
%
If the microshorts are too close to each other, trapping on the second
microshort does not happen (see references \cite{bp90pla,pb94pla} for 
details). We note that direct numerical simulations of the perturbed SG
equation (\ref{1}) (curve 4 of Figure \ref{1}) are in 
good correspondence with the trajectories of the system (\ref{2}). The
oscillations after the collision with the microshort can be attributed
to the fluxon radiation (not accounted by the first order perturbation
theory) and errors in determination of the fluxon center.

\begin{figure}[htb]
\centerline{\epsfig{file=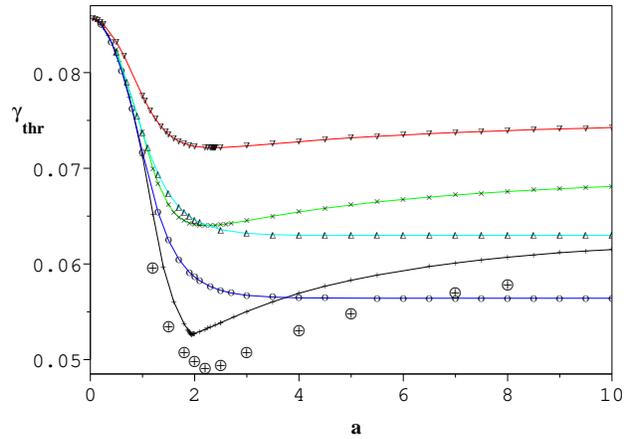,width=2.8in,angle=-90}}
\caption{Threshold pinning current as a function of the distance
between the microshorts for $\mu_1=0.6$, $\mu_2=0.4$ ($\Delta$); 
$\mu_1=\mu_2=0.5$ ($\circ$);
$\mu_1=0.4$, $\mu_2=0.6$ ($+$); $\mu_1=0.3$, $\mu_2=0.7$ ($\times$); 
$\mu_1=0.2$, $\mu_2=0.8$ ($\nabla$).
Results, obtained via direct numerical integration of the equations
of motion for $\mu_1=0.4$, $\mu_2=0.6$ are shown by $\oplus$.}
\label{fig2}
\end{figure}

The systematic evaluation of the threshold current $\gamma_{thr}$ as 
a function of the distance $a$ for different values of $\mu_1$ and
$\mu_2$ is shown in Figure \ref{fig2}. The resonant nature of the 
dependence of $\gamma_{thr}$ on $a$ for $\mu_1<\mu_2$ can be observed
clearly. While in the respective limiting cases it satisfies 
equation (\ref{6}), a resonant value  $a=a_r$ appears, at which the
threshold current attains its minimal value. 
%

The explanation of the resonant transmission can be done on the following 
qualitative argument. The analysis of the phase portraits in 
Figure \ref{1} shows that after being released from the microshort, 
the fluxon
accelerates in order to regain its equilibrium velocity $v_\infty$.
This
acceleration occurs in such a way that for some short interval the
fluxon
velocity exceeds the equilibrium value $v_\infty$. Therefore the fluxon 
has kinetic energy, which is larger than it
was while approaching the microshort from $X=-\infty$, and 
consequently it has enough energy to pass the microshort with the 
amplitude larger than $\mu_1$. Obviously, the best transmission would 
take place if $a$ slightly exceeds $|X_{2}-X_1|$. The 
estimation of the
resonant distance $a_r$ can be made from the analysis of the fluxon 
dynamics in the non-relativistic limit, given by
equations (\ref{6a})-(\ref{6b}). 
According to these equations the fluxon can be compared to the
particle that slides down along the potential
$U(X)=U(X \to \pm \infty) \sim - 2\pi \gamma X$. Depending on the value
of $\gamma$, it can be trapped in one of the wells of this potential
(shown in the inset of Figure \ref{fig1}). If the distance between 
microshorts is small enough, it can be considered as one microshort with
the renormalized
strength ${\bar \mu}(a)=\mu_1+\mu_2/\cosh^2 {a}$. The trapping can occur
at the only
existing minimum $X_1$ (see curve $1$ in the inset of
Figure \ref{fig1}) as shown
by the trajectory $1$ of Figure \ref{fig1}. As $a$ increases, 
the potential 
barriers separate and a local minimum $X_3$ appears, as shown by curves
$2-4$ in the inset of Figure \ref{fig1}.  If a new minimum appears, 
the trapping can 
occur also at the second microshort, as shown by the trajectory $2$ 
of Figure \ref{fig1}. Plots of the potential $U(X)$
clearly demonstrate that the shape of the barrier will be optimal 
when the minimum at $X=X_3$ is quite shallow. Since the half-width
of the function $\cosh^{-2}(X)$ is of the order of unity, 
it is expected that optimal separation of barriers occurs at $a \sim 2$. 
Numerical evaluation of $a_r$ confirms this estimate: 
$a_r=1.94$ (for $\mu_1=0.4$, $\mu_2=0.6$);  $a_r=2.24$ 
($\mu_1=0.3$, $\mu_2=0.7$) and $a_r=2.36$ (for $\mu_1=0.2$, $\mu_2=0.8$). 
If $\mu_1\ge\mu_2$, the transmission scenario is always 
determined by the first microshort and the trapping occurs only at 
$X=X_1$. Therefore the dependence $\gamma_{thr}$ on $a$ is
monotonically decreasing as shown in Figure \ref{fig2} 
for $\mu_1=0.6$, $\mu_2=0.4$.

\begin{figure}[htb]
\centerline{\epsfig{file=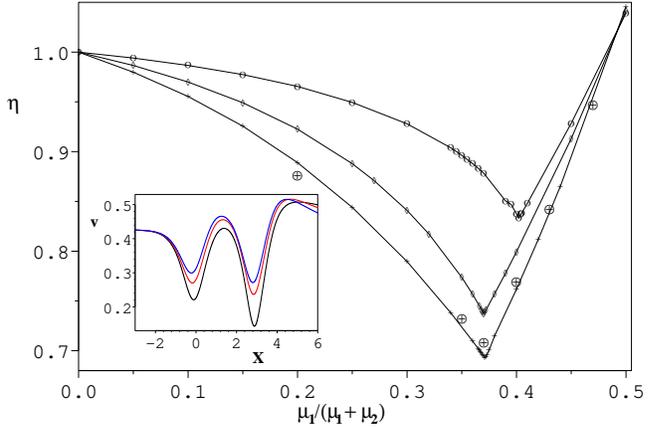,width=2.8in,angle=-90}}
\caption{Dependence of the enhancement factor (\ref{9})  on the
ratio $\mu_1/(\mu_1+\mu_2)$ $\alpha=0.1$ ($\circ$), $\alpha=0.2$ ($\diamond$)
and $\alpha=0.3$ ($+$). Solid lines are used as a guide for an eye. 
Results, obtained via direct numerical integration of the 
SG equation with the 4th order Runge-Kutta method
for $\alpha=0.3$ are shown by $\oplus$. The inset shows trajectories in the 
phase plane $(X,v)$ for $\alpha=0.1$, $\gamma=0.06$ (black line); 
$\alpha=0.2$, $\gamma=0.12$ (red line) and $\alpha=0.3$, $\gamma=0.18$
 (blue line). Other parameters are $\mu_1=0.4$, $\mu_2=0.6$, $a=2$.}
\label{fig3}
\end{figure}

It would be of interest to compare how the threshold pinning current 
depends on the dissipation parameter $\alpha$ and the ratio of the 
microshort amplitudes $\mu_1$ and $\mu_2$. Since the resonant
distance $a_r$ weakly depends on $\mu_{1,2}$ and
the pinning current depends strongly on the dissipation constant, it
is convenient to normalize $\gamma_{thr}(a,\mu_1,\mu_2;2)$ to the
pinning current on the strongest microshort \\
$\max{[\gamma_{thr}(\mu_1;1),\gamma_{thr}(\mu_2;1)]}$.
In Figure \ref{fig3} the dependence of the enhancement factor
\begin{equation}
\eta(a,\mu_1,\mu_2;2)=
\frac{\gamma_{thr}(a,\mu_1,\mu_2;2)}
{\max{[\gamma_{thr}(\mu_1;1),\gamma_{thr}(\mu_2;1)]}}~,
\label{9}
\end{equation}
on the ratio $\mu_1/(\mu_1+\mu_2)$ for different values of 
dissipation is shown. The value of the distance between the microshorts 
has been fixed to $a=2$. Increase of dissipation does not change 
much the resonant values of $\mu_{1,2}$. 
However, the value of the enhancement factor at the minimum 
decreases significantly. 
%
In the inset of Figure \ref{fig3} comparison of the fluxon slowing 
down on the microshorts is shown for different values of dissipation
and dc bias. Note that the ratio $\alpha/\gamma$ was kept constant in
order to fix the equilibrium velocity $v_\infty$. For stronger 
dissipation the 
fluxon slows down to smaller velocities (compare the black and 
blue curves that correspond to $\alpha=0.1$
and $\alpha=0.3$, respectively).  Therefore after release from
the microshort the fluxon can accelerate to greater values of velocity. 
As a result, it has more kinetic energy to pass the second microshort.
In other words, for larger dissipation one
needs larger bias, $\gamma$. Therefore, the tilt of the 
potential $U(X)$ increases and it smears out the
inhomogeneities, created by the impurities.

It should be emphasized 
that the validity of the perturbation theory approach has
 been confirmed
by the direct numerical integration of the original perturbed 
SG equation (\ref{1}).
In Figure \ref{fig2}, $\gamma_{thr}$ has been computed  via integration of
equation (\ref{1}) for $\mu_1=0.4$ and $\mu_2=0.6$. 
It is evident that the 
perturbation theory gives qualitatively the same result and the
quantitative difference is not very large. Similarly, 
in Figure \ref{fig3} the
results of the numerical integration 
of equation (\ref{1}) with $\alpha=0.3$ are 
given alongside with the perturbation theory results. A good 
qualitative and
quantitative correspondence between these two types of results is
clearly demonstrated. Therefore the usage of the approximation 
(\ref{2})-(\ref{3}) is justified.

\subsection{Transmission through $N>2$ microshorts}
\label{s3.2}

\begin{figure}[htb]
\centerline{\epsfig{file=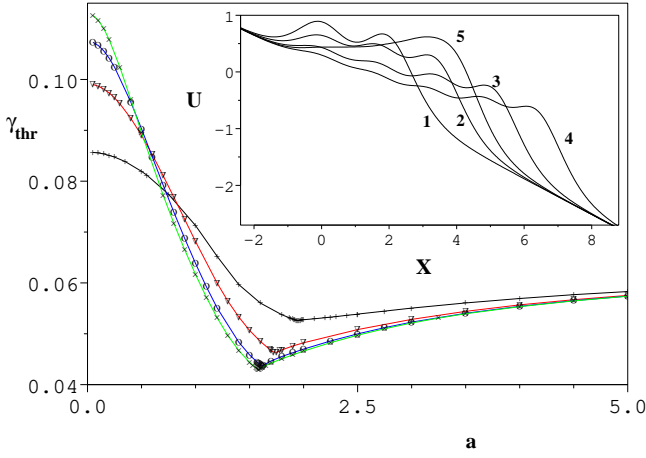,width=2.9in,angle=-90}}
\caption{
Threshold pinning current as a function of the distance
between the microshorts for 
$\alpha=0.1$, $N=2$, $\mu_1=0.4$, $\mu_2=0.6$ ($+$); 
$N=3$, $\mu_1=0.27$, $\mu_2=0.4$, $\mu_3=0.6$ ($\nabla$); 
$N=4$, $\mu_1=0.18$, $\mu_2=0.27$, $\mu_3=0.4$, $\mu_4=0.6$ ($\circ$) 
and 
$N=5$, $\mu_1=0.12$, $\mu_2=0.18$, $\mu_3=0.27$, $\mu_4=0.4$, $\mu_5=0.6$ 
($\times$). Solid lines are used as an guide for an eye.
The inset depicts the effective potential $U(X)$ [see 
equation (\ref{6b})]
for the configurations, described in the main figure at $\gamma=0.05$
and for the same values of $\{ \mu_n \}$:  $N=2$, $a=2$ (curve 1); 
$N=3$, $a=1.7$ (curve 2); $N=4$, $a=1.6$ (curve 3); 
$N=5$, $a=1.6$ (curve 4) and $a=1$ (curve 5). 
}
\label{fig4}
\end{figure}

Now we extend the results of the previous subsection on the case of
more then $N=2$ microshorts. In Figure \ref{fig4} the dependence of 
$\gamma_{thr}$ on $a$ for $N=2,3,4,5$ is presented. 
It clearly demonstrates that addition of an extra impurity to the left
from the weakest one decreases further the minimum of the threshold
current.    
%

The explanation can be easily seen with the help of the effective potential
$U(X)$ [see equation (\ref{6b})]. Its shape changes significantly when
extra microshorts are added. Comparing curves 1 and 2 in the inset of 
Figure \ref{fig4} one can see that the energy barrier, which the fluxon
should cross, lowers. Adding yet another microshort further lowers 
the barrier (see curves 3 and 4), so that in the interval $0<X<(N-1)a$
the potential barrier almost turns into the decaying slope which is
less steep then $-2 \pi \gamma X$. Decrease of $a$ leads to the gradual
raising of this slope (compare curves 4 and 5) and consequently to
increase of $\gamma_{thr}$.

\subsection{Transmission through microresistors}
\label{s3.3}

\begin{figure}[htb]
\centerline{\epsfig{file=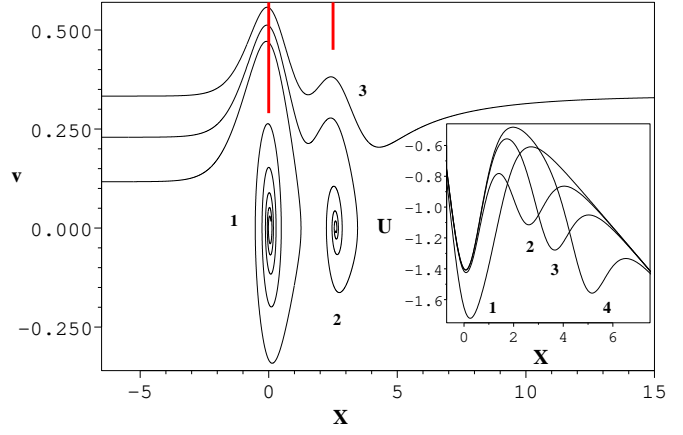,width=2.8in,angle=-90}}
\caption{
Phase space trajectories of equations (\ref{2})-(\ref{3}) 
for $N=2$ microresistors
with $\mu_1=-0.7$, $\mu_2=-0.3$ and distance $a=2.5$ between them. 
Curve $1$ corresponds to $\gamma=0.015$, curve $2$ - to $\gamma=0.03$
and curve $3$ -  to $\gamma=0.045$. Dissipation for all cases is 
$\alpha=0.1$. The inset shows the effective potential 
$U(X)$ [equation (\ref{6b})]
for $\mu_1=-0.7,\mu_2=-0.3$, $\gamma=0.03$, $a=1$ (curve 1),
$a=2.5$ (curve 2), $a=3.5$ (curve 3) and $a=5$ (curve 4).
Red vertical bars denote locations of the microresistors and the lengths
of the bars is proportional to their amplitudes, $\mu_{1,2}$.
}
\label{fig5}
\end{figure}

A microresistor is an attracting impurity, therefore the fluxon 
accelerates when approaching it and decelerates back to $v=v_\infty$
after passing through or remains trapped if its velocity (and 
consequently the external bias current) is less than the threshold 
value. The effective potential $U(X)$ for a microresistor corresponds
to the potential well. If two different microresistors are added 
consequently, the fluxon can be trapped on the first or on the second
one, or, if the bias is large enough, pass through. 
In Figure \ref{fig5} the phase portraits for the system with $N=2$ 
microresistors is shown. The change of the shape of $U(X)$ for 
the different distances between the microresistors is shown 
in the inset of Figure \ref{fig5}.
%

\begin{figure}[htb]
\centerline{\epsfig{file=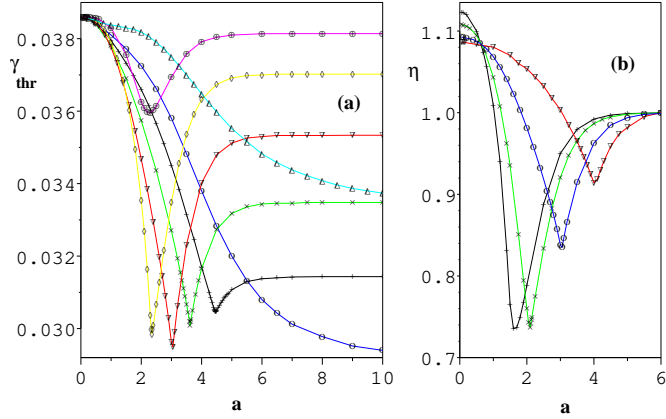,width=2.7in,angle=-90}}
\caption{
Panel (a).
Threshold pinning current as a function of the distance
between two microresistors for $\mu_1=-0.3$, $\mu_2=-0.7$ ($\Delta$); 
$\mu_1=\mu_2=-0.5$ ($\circ$);
$\mu_1=-0.6$, $\mu_2=-0.4$ ($+$); $\mu_1=-0.7$, $\mu_2=-0.3$ ($\times$); 
$\mu_1=-0.8$, $\mu_2=-0.2$ ($\nabla$); $\mu_1=-0.9$, 
$\mu_2=-0.1$ ($\diamond$)
and $\mu_1=-0.97$, $\mu_2=-0.03$ ($\oplus$). Damping coefficient 
equals $\alpha=0.1$.
Panel (b). Enhancement factor $\eta$ [see equation (\ref{9})] as
a function of $a$ 
for $\mu_1=-0.8$, $\mu_2=-0.2$ and different values of damping $\alpha$:
$\alpha=0.3$ ($+$), $\alpha=0.2$ ($\times$), $\alpha=0.1$ ($\circ$)
 and $\alpha=0.05$ ($\nabla$). }
\label{fig6}
\end{figure}

The computation of the threshold current $\gamma_{thr}$ shows that 
resonant fluxon transmission is possible if $\mu_1 < \mu_2$ and 
does not happen if $\mu_1 \ge \mu_2$ (see Figure \ref{fig6}a). 
Explanation of this phenomenon is similar to the case of two 
microshorts. If the microresistors are located very close to each other,
then their amplitudes add up and the fluxon interacts with the 
microresistor of the amplitude $\mu \simeq \mu_1+\mu_2$. When the 
impurities start to separate, the effective energy barrier which
the fluxon should surmount, lowers (compare curve $1$ with curves $2$
and $3$ in the inset Figure \ref{fig6}). The distance  
between the wells becomes optimal for the best fluxon transmission 
before they are completely separated (compare curves $3$ and $4$). 

%
In contrast to the transmission through two microshorts, the resonant 
value $a=a_r$ depends strongly on the amplitudes $\mu_{1,2}$. Indeed,
for $\mu_1=-0.7$, $\mu_2=-0.3$ one obtains $a_r\simeq 3.62$ and
for $\mu_1=-0.9$, $\mu_2=-0.1$ the resonant distance equals
$a_r\simeq 2.35$. Comparing curves $2-4$ in the inset of 
Figure \ref{fig5}, one can notice that the fluxon needs enough kinetic
energy to overcome the second maximum, located at $X=X_4$.
Obviously, if $X_2$ and $X_4$ are not enough separated, the fluxon will
have no time to accelerate in order to avoid trapping on the 
second microresistor. Therefore the case of curve $3$ is the most
optimal one: the height of the barrier at $X_2$ is not too large,
as compared to the curve $4$ and the distance between $X_2$ and $X_3$
is enough to gain velocity, sufficient for the successful passage
over the second barrier. These considerations, of course, 
correspond to the situation, when the impurities are not 
strongly separated.
Otherwise only the interaction with the first one would matter.   
For the same reason the position of the minimal threshold current, $a_r$,
(see Figure \ref{fig6}b) increases with decrease of the damping parameter
$\alpha$. Depth of the minimum
decays with decrease of $\alpha$ similarly to the
case of two microshorts because with the stronger bias the fluxon
can pass through the impurities much easier.

Putting additional microresistors after $\mu_2$, 
$0>\mu_N> \cdots >\mu_2>\mu_1$ further lowers the critical pinning
current similarly to the case of $N>2$ microshorts, described
in the previous Subsection.

\subsection{Transmission through a microshort and a microresistor}
\label{s3.4}

\begin{figure}[htb]
\centerline{\epsfig{file=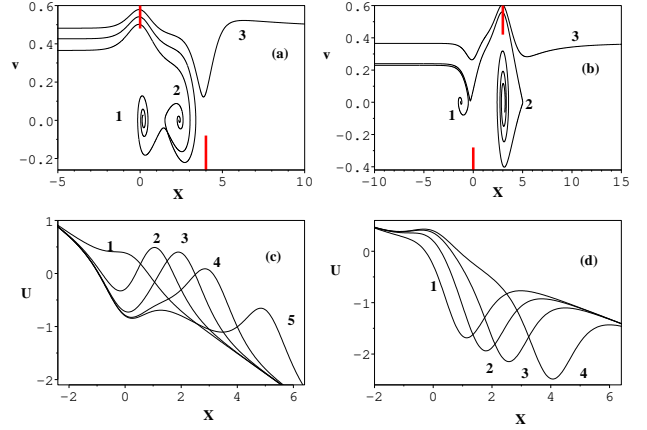,width=2.5in,angle=-90}}
\caption{
Phase portraits of the fluxon dynamics and effective potentials
for the fluxon dynamics in the case of one microshort and
one microresistor.
Panel (a) shows fluxon trajectories on the phase plane $(X,v)$
for $\alpha=0.1$, $\mu_1=-0.4$, $\mu_2=0.6$, $a=4$; $\gamma=0.05$ (curve 1),
$\gamma=0.06$ (curve 2) and $\gamma=0.07$ (curve 3).
Panel (b) shows fluxon trajectories on the phase plane $(X,v)$
for $\alpha=0.1$, $\mu_1=0.2$, $\mu_2=-0.8$, $a=3$; $\gamma=0.03$ (curve 1),
$\gamma=0.0314$ (curve 2) and $\gamma=0.05$ (curve 3).
Panel (c) shows the effective potential $U(X)$ for 
$\mu_1=-0.4$, $\mu_2=0.6$ and different values of $a$: 
$a=0.1$ (curve 1), $a=1$ (curve 2), $a=2$ (curve 3), $a=3$ (curve 4) and 
$a=5$ (curve 5).
Panel (d) shows the effective potential $U(X)$ for 
$\mu_1=0.2$, $\mu_2=-0.8$ and different values of $a$: 
$a=1$ (curve 1), $a=1.7$ (curve 2), $a=2.5$ (curve 3) and $a=4$ (curve 4).  
}
\label{fig7}
\end{figure}

\begin{figure}[htb]
\centerline{\epsfig{file=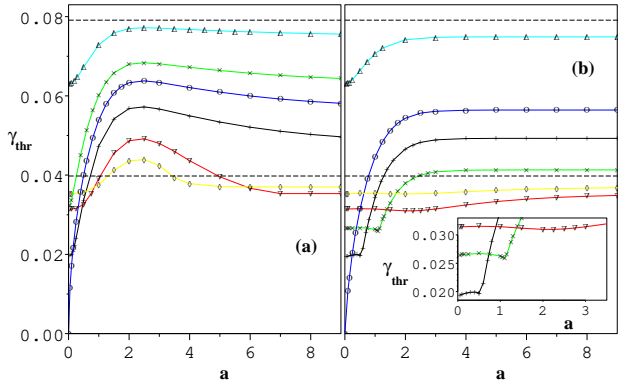,width=2.5in,angle=-90}}
\caption{
Threshold pinning current as a function of the distance
between two impurities $\mu_1 \mu_2 <0$ with $\alpha=0.1$.
Panel (a): $\mu_1=-0.2$, $\mu_2=0.8$ ($\Delta$); 
    $\mu_1=-0.4$, $\mu_2=0.6$ ($\times$); $\mu_1=-0.5$, $\mu_2=0.5$ ($\circ$);
    $\mu_1=-0.6$, $\mu_2=0.4$ ($+$); $\mu_1=-0.8$, $\mu_2=0.2$ ($\nabla$)
    and $\mu_1=-0.9$, $\mu_2=0.1$ ($\diamond$). 
Panel (b): $\mu_1=0.8$, $\mu_2=-0.2$ ($\Delta$); $\mu_1=0.5$, $\mu_2=-0.5$ ($\circ$);
     $\mu_1=0.4$, $\mu_2=-0.6$ ($+$); $\mu_1=0.3$, $\mu_2=-0.7$ ($\times$);
     $\mu_1=0.2$, $\mu_2=-0.8$ ($\nabla$) and $\mu_1=0.1$, $\mu_2=-0.9$ ($\diamond$).
Dashed lines correspond to the values $\gamma_{thr}(\pm 1; 1)$. 
Inset shows more detailed picture of $\gamma_{thr}=\gamma_{thr}(a)$.
}
\label{fig8}
\end{figure}

Finally, we consider the case when two impurities of different polarity 
(a microshort and a microresistor) are placed one after another.
If the microresistor is located before the 
microshort ($\mu_1<0, \mu_2>0$) resonant enhancement
of the threshold pinning current does not happen. In Figure \ref{fig7}
(panel a) the phase portraits for this case are shown. The microresistor
is an attracting impurity after which the fluxon slows down. On contrary,
the microshort is a repelling impurity and the fluxon slows down when 
approaching it. Therefore it is obvious that by placing impurities in such
a way one increases $\gamma_{thr}$ as compared to the case of each
individual impurity. The analysis of the effective potential $U(X)$,
shown in Figure \ref{fig7} (panel c), further confirms above considerations.
The height of the effective barrier, which the fluxon should overcome,
can be greater than the height of the individual barriers, created by the
individual impurities. If the impurities are very close their 
influences cancel
each other and the fluxon interacts with the impurity of the strength
$-|\mu_1|+\mu_2$. The dependence of the threshold pinning 
current $\gamma_{thr}$
on the distance between the impurities is given 
in Figure \ref{fig8} (panel a).
%
For $\mu_1=-\mu_2=-0.5$ the microshort and microresistor cancel each
other for $a=0$, therefore the dependence starts at zero and increases
until it reaches the maximal value and then decreases, tending 
monotonically to the value of $\gamma_{thr}$ that corresponds to one 
microshort with $\mu=0.5$. The dependence  of $\gamma_{thr}$ on $a$
shows an ``antiresonant'' behaviour because it has a maximum 
at some certain value of $a$. Analysis of the shape of $U(X)$ from 
the Figure \ref{fig7}c predicts that the worst transmission would occur
when the potential well and the barrier, created by the microresistor
and the microshort, respectively, separate
from each other far enough to create the highest total 
barrier (see the curve $3$ of Figure \ref{fig7}c), but not too far (as for the 
curve $5$ of the same figure) so that each impurity interacts 
individually with the fluxon.

Consider now the case $\mu_1>0, \mu_2<0$. The phase portraits for the 
fluxon dynamics are shown in Figure \ref{fig7} (panel b). The dependence 
of the threshold pinning current on the distance between the 
impurities is shown in Figure \ref{fig8} (panel b). In the case 
$\mu_1=-\mu_2=0.5$ at $a=0$ the impurities cancel each other. When
$a$ increases, $\gamma_{thr}$ monotonically increases, tending to the
threshold value of one isolated microshort with the amplitude $\mu=0.5$.
In this case trapping occurs only on the microshort because 
analysis of equations (\ref{4a})-(\ref{4b}) shows that
$\gamma_{thr}(0.5;1)>\gamma_{thr}(-0.5;1)$. 
If $\mu_1 > |\mu_2|$ the dependence of $\gamma_{thr}$ on $a$ is also
monotonic. At $a \simeq 0$ the fluxon ``feels'' both impurities as one
microshort with $\mu_1-|\mu_2|$. When the impurities separate, the contribution
of the microresistor to the total amplitude weakens and the threshold
current gradually increases till the value $\gamma_{thr}(\mu_1;1)$.
%
If $\mu_1$ decreases and $|\mu_2|$ increases, behaviour of the
critical pinning current on $a$ becomes more complicated. Consider 
first the case $\mu_1=0.4$, $\mu_2=-0.6$. In the neighbourhood of 
$a=0$ the system can be considered as a microresistor with the 
amplitude $\mu_1-|\mu_2|$. When $a$
increases, the well and the barrier, created by the microshort $\mu_1$
start to separate, increasing the depth of the well (created by the
microresistor). 
After some value of the distance $a$ the dependence of 
$\gamma_{thr}=\gamma_{thr}(a)$ 
experiences sharp breaking and $\gamma_{thr}$ starts to grow 
with $a$. 
The difference between trapping before this breaking point 
and after it is
based on the trapping scenario at $\gamma \le \gamma_{thr}$. 
For values of $a$ below the breaking point trapping occurs in the well, 
created by the microresistor
(curve $2$ of Figure \ref{fig7}b) while after the breaking 
point trapping occurs
on the microshort. In other words, the breaking point 
signals the value
of the separation of the impurities, before which the fluxon ``feels''
them as one microresistor and after which the fluxon ``feels'' 
them separately.
Decrease of $\mu_1$, and subsequent increase of $|\mu_2|$ 
leads to the 
gradual shift of the breaking point to the right and
smoothing of the shape of the dependence $\gamma_{thr}(a)$.
Further decrease of $\mu_1$ and 
increase of $|\mu_2|$ makes the dependence $\gamma_{thr}(a)$ more
and more flat, so that in the limit $\mu_1 \to 1$, $\mu_2 \to 0$
it tends to the horizontal line $\gamma_{thr}=\gamma_{thr}(-1;1)$.

\section{Conclusions}
\label{s4}

We have investigated the fluxon transmission in a dc-biased long Josephson 
junction (LJJ) through two or more impurities of
different polarity: microshorts and/or microresistors. We have observed
that the threshold pinning current can depend on the distance between
impurities in the resonant way for the case of two or more microshorts
or two or more microresistors. That means that at some value
of the distance the threshold current attains a minimal value,
which is less than the threshold current of the strongest impurity.
The resonant transmission does not occur if the fluxon interacts with
two impurities of different sign: a microshort and a microresistor.

The observed effect should not be confused with the resonant soliton 
transmission
in the non-dissipative cases \cite{ksckv92jpa,zkv94jpsj}.
In the case of fluxon dynamics in a long Josephson junction the presence
of dissipation is unavoidable. Far away from the impurities fluxon exists
as an only one attractor of the system with the velocity, predefined by the
damping parameter and external bias. Therefore, contrary to the 
non-dissipative case, there is no sense in computing the transmission
ratio, which in 
our case can take only two values: zero (trapping) and unity (transmission).
Also it should not be confused with the fluxon tunneling
as a quantum-mechanical object \cite{sb-jm97prb} 
across the double-barrier potential, created by two identical
microshorts. 

The discussed phenomenon can be observed experimentally in an annular
LJJ via monitoring the current-voltage (IV) characteristics. 
For a LJJ with one impurity the fluxon IV curve has a hysteresis-like
nature with two critical values of the dc bias (discussed in 
Section \ref{s2}).
The lower one is the threshold pinning current, which is the smallest
current for which fluxon can propagate. Although simulation in this
paper have been performed for the infinite junction, there should be
no principal differences with the case of an annular junction with 
sufficiently large length $L\gg \lambda_J$. Currently experiments are
performed in the junctions $L \sim 10\lambda_J$ (see \cite{u98pd}) that
can be considered as {\it long}.

For the future research in this direction it is of interest to
find out how the resonant fluxon transmission changes if the actual 
size of the impurities and the junction width along the
$y$-axis are taken into account.

\section*{Acknowledgments}
This work has been supported by DFFD, project number GP/F13/088, 
and the special program of the National Academy of Sciences of Ukraine
for young scientists.


\begin{thebibliography}{10}
\bibitem{parameters} Expressions for the Josephson penetration depth
and Josephson plasma frequency and their actual values 
can be found in \cite{barone82}.
\end{thebibliography}
\end{document}